%% file: ma.tex
\documentclass[review]{elsarticle}
\makeatletter
\def\ps@pprintTitle{%
 \let\@oddhead\@empty
 \let\@evenhead\@empty
 \def\@oddfoot{\centerline{\thepage}}%
 \let\@evenfoot\@oddfoot}
\makeatother

\usepackage[utf8]{inputenc} 
\usepackage[T1]{fontenc}    
\usepackage{lmodern}
\usepackage{hyperref}       
\usepackage{url}            
\usepackage{booktabs}       
\usepackage{amsfonts}       
\usepackage{nicefrac}       
\usepackage{microtype}      
\usepackage{lipsum}
\usepackage{array}
\usepackage{threeparttable}
\usepackage{nomencl}
\usepackage{graphicx}
\usepackage{multirow}
\usepackage{amsmath}
\usepackage{epstopdf}
\usepackage{rotating}

\usepackage[margin=3cm]{geometry}

\newcolumntype{L}[1]{>{\raggedright\let\newline\\\arraybackslash\hspace{0pt}}m{#1}}
\newcolumntype{C}[1]{>{\centering\let\newline\\\arraybackslash\hspace{0pt}}m{#1}}
\newcolumntype{R}[1]{>{\raggedleft\let\newline\\\arraybackslash\hspace{0pt}}m{#1}}

\input{symbols.tex}

\begin{document}

\title{Challenges and opportunities of inertia estimation and forecasting in low-inertia power systems}
\begin{frontmatter}
\author{
	Heylen,~E.,  Strbac,~G and Teng,~F\textsuperscript{*}\\
    Control and Power Research Group\\
	Department of Electrical and Electronic Engineering\\
	Imperial College London\\
	*Correspondence: f.teng@imperial.ac.uk}

\begin{abstract}
Accurate inertia estimates and forecasts are crucial to support the system operation in future low-inertia power systems. A large literature on inertia estimation methods is available. This paper aims to provide an overview and classification of inertia estimation methods. The classification considers the time horizon the methods are applicable to, i.e., offline post mortem, online real time and forecasting methods, and the scope of the inertia estimation, e.g., system-wide, regional, generation, demand, individual resource. Shortcomings of the existing inertia estimation methods have been identified and suggestions for future work have been made.
\end{abstract}

\end{frontmatter}
Date: 21/08/2020

\input{intro2.tex}

\input{operations.tex}

\input{classification.tex}
\input{conclusion2.tex}

\section*{Acknowledgments}
This work has been funded by Electricity Network Innovation Allowance project on Short-term System Inertia Forecast NIA\_NGSO0020.

\bibliographystyle{IEEEtran}  
\bibliography{references2}  

\end{document}

%% file: symbols.tex
\newcommand{\group}{g}
\nomenclature[A]{$\group$}{Type of generators}
\newcommand{\gen}{i}
\nomenclature[A]{$\gen$}{Generators}

\newcommand{\Group}{\mathcal{G}}
\nomenclature[B]{$\Group$}{Types of generators}

\newcommand{\Hconstant}[2]{H^{#1}_{#2}}

\newcommand{\Energyh}[2]{\hat{E}^{#1}_{#2}}
\newcommand{\Energyb}[2]{\bar{E}^{#1}_{#2}}
\newcommand{\Hhconstant}[2]{\hat{H}^{#1}_{#2}}
\newcommand{\Hbconstant}[2]{\bar{H}^{#1}_{#2}}
\newcommand{\Havg}{\Hbconstant{avg}{}}
\nomenclature[C]{$\Havg$}{Estimated average inertia constant over all generators}
\newcommand{\Hsync}{\Hhconstant{sync}{}}
\nomenclature[C]{$\Hsync$}{Forecasted inertia constant from synchronous generation}

\nomenclature[C]{$\Hsync$}{Equivalent inertia constant for the total system}

\nomenclature[C]{$\Hsync$}{Forecasted kinetic energy from synchronous generation}
\newcommand{\Hgroup}{\Hbconstant{}{\group}}
\nomenclature[C]{$\Hgroup$}{Estimated average inertia constant for a certain type of generators}
\newcommand{\Hgen}{\Hconstant{}{\gen}}
\nomenclature[C]{$\Hgen$}{Exact inertia constant for an individual generator}
\newcommand{\Hbgen}{\Hbconstant{}{\gen}}
\nomenclature[C]{$\Hbgen$}{Estimated inertia constant for an individual generator}
\newcommand{\Esys}{\Energyh{sys}{}}
\nomenclature[C]{$\Esys$}{Forecasted total system kinetic energy}
\newcommand{\Edemand}{\Energyb{D}{}}
\nomenclature[C]{$\Edemand$}{Estimated kinetic energy from demand}
\newcommand{\Cgen}{\Power{C}{\gen}}
\nomenclature[C]{$\Cgen$}{Capacity of generator $\gen$}

\newcommand{\Power}[2]{P^{#1}_{#2}}
\newcommand{\Powerh}[2]{\hat{P}^{#1}_{#2}}
\newcommand{\Pgen}{\Powerh{G}{\gen}}
\nomenclature[D]{$\Pgen$}{Forecasted online capacity per individual generator $\gen$}
\newcommand{\Pgroup}{\Powerh{G}{\group}}
\nomenclature[D]{$\Pgroup$}{Forecasted online capacity per type of generator $\group$}
\newcommand{\Ptot}{\Powerh{Tot}{}}
\nomenclature[D]{$\Ptot$}{Forecasted total online capacity }

\newcommand{\PN}{PN}

\nomenclature[D]{$\PN$}{Forecasted status of generator $\gen$}

\newcommand{\Sbase}{S^{base}}
\newcommand{\freq}{f}
\newcommand{\freqpu}{\bar{f}}

\newcommand{\fdiffpu}{\Delta \freqpu}
\newcommand{\ti}{t}
\newcommand{\damping}{K^{D}}
\newcommand{\Pdiff}{\Delta P}

%% file: intro2.tex
\section{Introduction}
\label{sec:intro}
Sufficient inertia in power systems is crucial to ensure secure and reliable power supply. In traditional power systems, inertia was provided by the rotating masses of synchronous generation and motor loads. These days, inertia is decreasing due to the massive introduction of converter-interfaced generation, consumption and transmission of power, which causes challenges to maintain acceptable frequency profiles. Inertia has been specifically low during periods of low demand and high renewable production \cite{national_grid_eso_response_2019}. To deal with the challenges of reduced inertia levels, system operators would benefit from accurate estimates or measurements of the inertia available in the system at each point in time as well as from forecasts of expected inertia levels \cite{Matevosyan_implementation_2018}. 
System operators currently estimate system inertia based on part of the resource \cite{Matevosyan_implementation_2018} or based on a simplified empirical relation \cite{wilson_measuring_2019}, which are approximate methods. Considering the evolution of power systems towards lower inertia levels, reliable and more accurate methods to estimate instantaneous and forecast future inertia are crucial.

Although a large literature on inertia estimation methods is available, they have not been surveyed so far. Existing review papers in the field of low-inertia power systems have focused on (i) the role and control of technical inertia enhancement techniques involving converter-interfaced energy buffers, also denoted as virtual inertia or inertia emulation \cite{tamrakar_virtual_2017,dreidy_inertia_2017,fang_inertia_2019,fernandez-guillamon_power_2019,ratnam_future_2020}, (ii) a general overview of the relevance of inertia in power systems by looking at the different sources of inertia and their impact on the operation of power systems \cite{tielens_relevance_2016} and (iii) a broad survey of the issues related to low-inertia power systems as well as the solutions that have been put forward with a particular focus on transient and frequency stability and converter control \cite{milano_foundations_2018}. 

To give insight in the shortcomings, challenges and opportunities of inertia estimation methods in the operation of low-inertia power systems, this paper presents an overview and classification of available inertia estimation methods. We surveyed the literature on inertia estimation methods and categorized the approaches according to the applicable time horizon and the scope of the inertia estimates. This enables us to discuss the evolution in the development of inertia estimation methods with relation to the operational practice and to identify shortcomings of the existing methods. Pathways for future research in the field of inertia estimation methods have been put forward.


%% file: operations.tex
\section{Operational practice with respect to inertia}
\label{sec:operations}
Secure and reliable operation of power systems requires a frequency that is nearly constant around the nominal frequency, which equals 50Hz in most parts of the world and 60Hz in the Americas and some parts of Asia. Grid codes prescribe acceptable frequency deviations and the system operator should take adequate actions to satisfy these limits. Inertia is a crucial parameter to satisfy the frequency limits in today's power systems.

\subsection{Operational constraints}
\label{sec:operationalconstraints}
\begin{figure}[htbp]
	\centering
	\includegraphics[width=\linewidth]{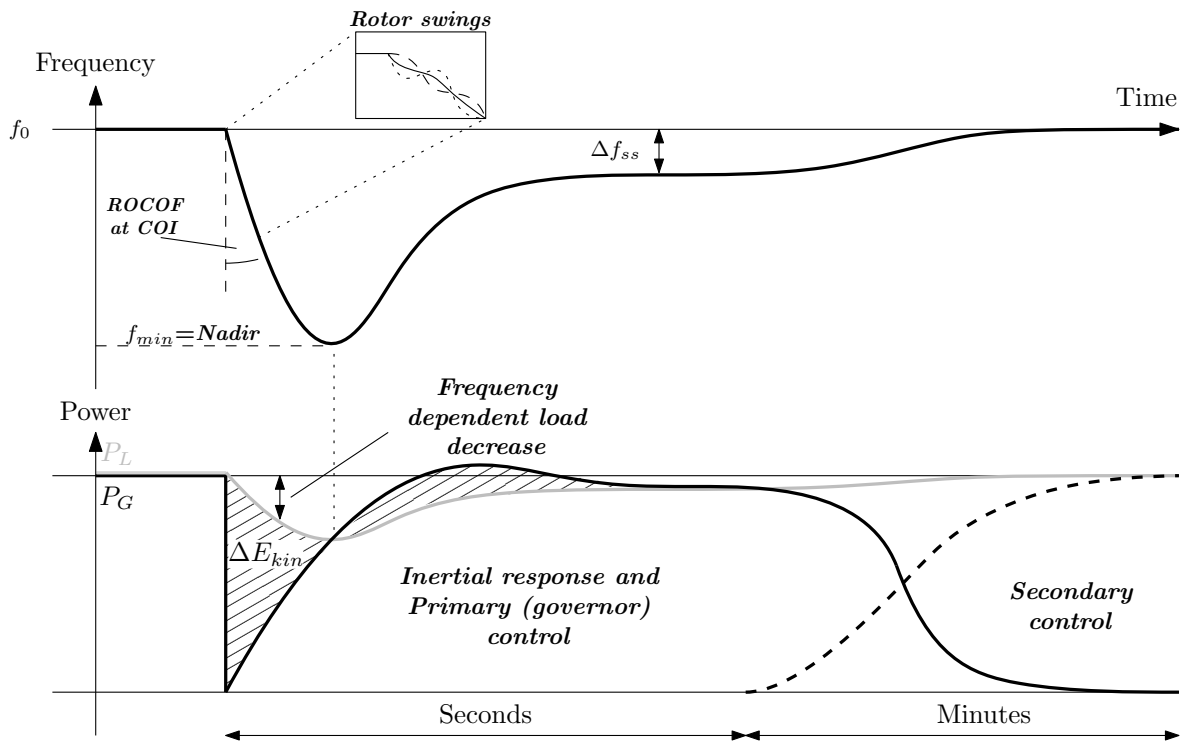}
	\caption{Frequency response after a power imbalance [Figure from \cite{tielens_operation_2017}, used with permission from the author]}
	\label{fig:f_response}
\end{figure}

The frequency deviates due to imbalances between demand and supply as characterized by the swing equation:\footnote{This is the linearised version of the swing equation, which is valid for normal oscillations in power systems \cite{anderson_dynamics_2012}.}
\begin{equation}
2\cdot H \cdot \frac{d \fdiffpu}{d\ti} + \damping \cdot \fdiffpu = \frac{\Pdiff}{\Sbase} \label{eq:swing}
\end{equation}
Where $H$ is the system inertia constant [s], 
$\fdiffpu$ the relative frequency deviation from the nominal frequency [pu], $\damping$ the damping coefficient [\%], i.e., the percent change in system load per percent change in grid frequency (This constant varies between 1-3\% across different power systems and is about 2.5\% for the Great Britain system \cite{taylor_forecasting_2016}) and $\Sbase$ the base power [MVA]. Figure \ref{fig:f_response} shows a typical frequency response after a power imbalance.\footnote{The frequency response in Fig. \ref{fig:f_response} gives the frequency evolution at the centre of inertia (COI). The frequency at the centre of inertia is a theoretical concept representing the inertia weighted frequency of all generators in the system and cannot be measured directly in the system. Calculating this value would require individual frequency measurements for all generators. Instead of estimating the frequency at the COI online, system operators typically measure the frequency at a relevant pilot node in the system. However, frequency measurements should be done with care, as they are heavily influenced by the location of the measurement and neighbouring generators \cite{milano_foundations_2018}.}

Grid codes put limits on the rate of change of frequency (ROCOF) $\frac{d \fdiffpu}{d\ti}$ and the frequency nadir $\freq_{min}$ to avoid tripping of protection relays. Generating units' protection relays are configured to avoid generating units to be exposed to ROCOF ranges from 1.5 - 2 Hz/s (over a 500-ms rolling window) that can cause pole slipping and catastrophic failure \cite{eirgrid/soni_rocofindependent_2013}. Also anti-islanding protection may be designed based on the detection of ROCOF \cite{operator_international_2016}. Frequency nadir is limited to avoid tripping of underfrequency load shedding protection relays. Fang et al. provide an overview of the ROCOF and frequency nadir limits in different power systems around the world \cite{fang_inertia_2019}.

Power system inertia is an important instrument to limit the ROCOF and frequency nadir. In its elementary form, inertia is defined as the resistance of a physical object to a change in its state of motion, including changes in speed and direction \cite{tielens_operation_2017}. In a power system context, inertia can be understood as the resistance, in the form of any kind of energy exchange, to counteract the changes in system frequency resulting from power imbalances in generation and demand \cite{tielens_operation_2017}. In the Great Britain system, the inertial response is defined as the period up to one second immediately following a loss of generation or demand prior to the activation of primary frequency response services \cite{ashton_inertia_2015}. The inertial response combined with the droop response of the primary frequency control limits the frequency nadir after a disturbance. The frequency restoration is further accomplished by the secondary and tertiary frequency control. Rebours et al. provide an overview of the technical features of the primary, secondary and tertiary frequency control reserves in different power systems \cite{rebours_survey_2007}.   

\subsection{Contributors to the power system inertia}
\label{sec:contributors}
The nature and size of the available inertial response in the power system has been changing over the last decade. Originally, rotating masses of synchronous machines, both generation and load, exchange kinetic energy with the system in the case of a power imbalance by slowing down or speeding up. This impacts the system frequency as the rotor speed is proportional to the frequency in the system. The inertial response of the synchronous rotating masses is instantaneous. 
In today's power system, synchronous machines have been gradually replaced by converter-integrated generators and demand. Moreover, power-electronic-interfaced high voltage dc technology has been increasingly used for interconnectors between power systems. These technologies can offer inertial response if they have an adequate energy buffer and control system in place. The control system should regulate the release and absorption of energy similar to the inertial response of synchronous generation units. Fang et al. give an overview of the emerging techniques for inertia emulation with different sources of energy storage \cite{fang_inertia_2019}. The penetration of inertia emulation in power systems has been limited so far. First of all, appropriate regulation has not yet been in place. Second, the system's dynamics characteristics will be impacted as control characteristics of inertia emulation techniques differ compared to synchronous generators' behaviour in case of disturbances \cite{milano_foundations_2018, ackermann_paving_2017}. Third, emulated inertial response typically comes from more variable and uncertain energy buffers, which increases uncertainty and variability in system operation \cite{ruttledge_frequency_2012,noauthor_enhanced_2019}. This evolution implies that traditional, instantaneous inertial response in modern power systems has been 
declining and tends to reduce even further in the future.\footnote{In the extreme case of 100\% non-synchronous systems, natural response to power imbalances is no longer present in the system and frequency is no longer a measure of power imbalance \cite{milano_foundations_2018}. Although the viability of control and operation of zero-inertia power systems have been proven in simulation studies and practical experiments, their practical implementation will not be for tomorrow, as many existing grid-connected devices and machines expect frequency to change relatively slowly and further study is required to integrate new converter technology into existing power system structures \cite{ackermann_paving_2017}. Milano et al. expect that substituting all conventional power plants with converter-interfaced generation will take a few decades \cite{milano_foundations_2018}.}

\subsection{Operational measures to limit the risk of ROCOF and underfrequency relay tripping}
\label{sec:actions}

\begin{figure}[htbp]
	\centering
	\includegraphics[width=0.9\linewidth]{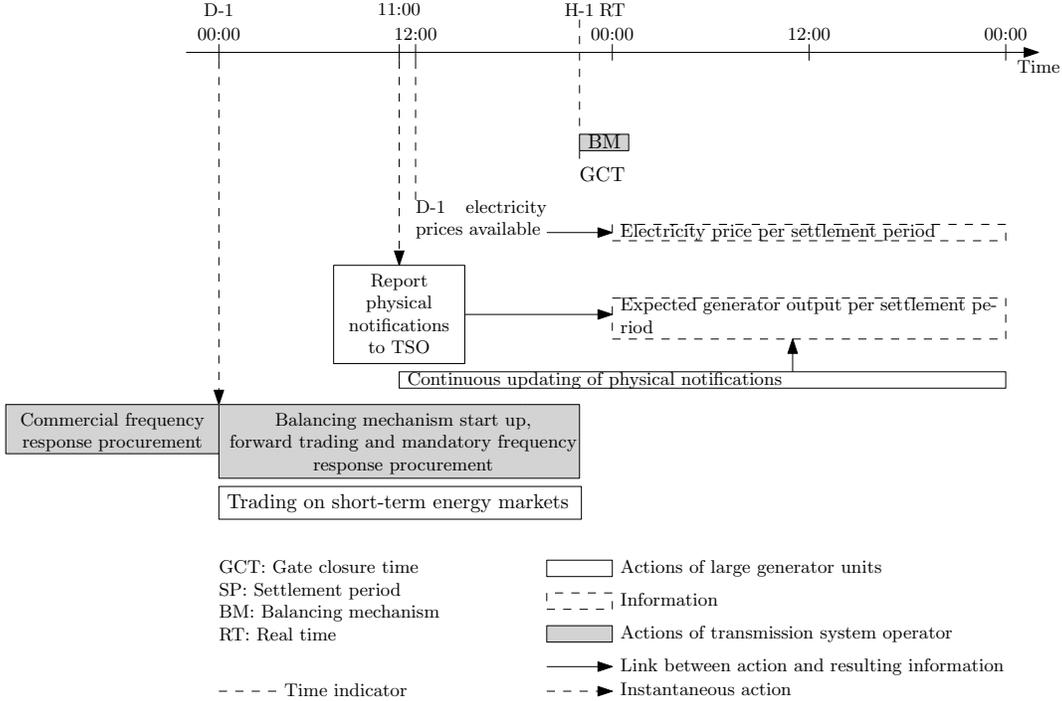}
	\caption{Overview of the operational practice related to inertia for the case of Great Britain}
	\label{fig:TSO_procedure}
\end{figure}

Reduced instantaneous inertial response increases the risk of tripping loss of main protections of embedded generators and underfrequency load-shedding relays. Besides changing grid codes in terms of connection criteria for non-synchronous generation \cite{brisebois_wind_2011} or modifying protection relay settings to cope with higher ROCOF and larger frequency swings \cite{bradley_dc0079_2017}, operational measures can be taken to reduce this risk. 
Considering the different terms in the swing equation (Eq. \eqref{eq:swing}), the system operator can reduce the risk of tripping ROCOF protection relays by reducing the possible largest loss in the system $\Pdiff^{max}$ or by increasing the inertial response, represented by the inertia constant $H$. Fig. \ref{fig:TSO_procedure} gives an overview of the operational practice with respect to inertia for the case of Great Britain.

Reducing the possible largest loss requires the system operator to redispatch generation or interconnection flows ahead of real time. In Great Britain, the system operator is informed by the expected output of the balancing mechanism units, which include the largest generating units, from 11:00 day ahead for each 30-minute settlement period of the next day via the physical notifications. The grid code of Great Britain defines physical notifications as "a series of MW figures and associated times, making up a profile of intended input or output of active power at the grid entry point or grid supply point, as appropriate" \cite{noauthor_balancing_nodate}. Large generator units, interconnectors and large demand units should provide these physical notifications. They should update their physical notifications throughout the day if significant changes are to be expected \cite{noauthor_balancing_nodate}. One hour ahead of real time, the final market settlement is reported to the system operator and the balancing mechanism starts. During the balancing mechanism, the system operator can accept bids and offers of the balancing mechanism units and request them to increase or decrease their power output. System operator's decisions upon generator redispatch are not only affected by economic concerns, but should also respect generators' reported ramp rate limits, minimum run time and the time taken to synchronize, as well as network constraints while being practicable for the system operator, e.g., large volumes of energy are preferred to reduce the number of instructions. Nowadays, reducing the largest loss is typically 10 times cheaper than increasing the inertial response by bringing additional generators online.\footnote{\url{https://www.nationalgrideso.com/document/152351/download} Accessed 03/03/2020} However, its effectiveness in power systems with a reduced number of large synchronous generator units and more smaller, distributed generation units, which are highly variable and uncertain in nature, is questionable.

To increase system inertia, the system operator can replace large amounts of converter-interfaced generation with synchronous generation ahead of real time. If additional synchronous generators need to be brought online, the system operator should take into account the time to synchronize the power plant. To deal with long synchronization time, the system operator has the possibility to start up balancing mechanism units that are not expected to run according to the physical notifications via the balancing mechanism start up ancillary service at the price of a start-up and hot-standby payment \cite{noauthor_proposed_2006} or using forward trading. These actions can be initiated before the balancing mechanism starts. 

On a time scale of seconds, frequency response can prevent the tripping of underfrequency-load-shedding relays. The system operator should adequately contract and procure frequency response services ahead of real time. Moreover, to deal with reduced inertia levels causing increased ROCOF and larger frequency swings, system operators are developing new, faster-acting frequency response services to ensure frequency containment within operational limits and avoid underfrequency load shedding with relaxed ROCOF limits \cite{national_grid_eso_response_2019}.\footnote{National Grid is currently in the process of replacing the traditional primary, secondary and tertiary frequency response services with a new integrated suite of services \cite{national_grid_eso_response_2019}. }
Alternatively, markets for inertia, which enable the trade of virtual inertia from converter-interfaced resources, have been suggested in the literature \cite{ela_market_2014}. It has been shown that inertia is especially a valuable product if it reduces wind curtailment in the system \cite{badesa_economic_2017}. 

Nowadays, National Grid ESO builds tables representing the five-minute resolution frequency response requirement as a function of demand, inertia and largest infeed loss using a frequency simulation engine \cite{national_grid_eso_appendices_nodate}. These tables are used in operational systems to optimise the overall frequency response requirement for 5 minute blocks throughout the day based on the real-time demand, inertia and largest infeed loss. Combining these requirements with the contractual arrangements per settlement period, gives the control room engineer insight in the residual frequency response requirements that should be procured from mandatory services in real time.\footnote{\url{https://www.nationalgrideso.com/document/152351/download} Accessed 03/03/2020} Control methods to coordinate the delivery of multiple frequency response services in real time have been investigated in \cite{noauthor_enhanced_2019}. In power systems operated by independent system operators, stochastic unit commitment models for simultaneous scheduling of multiple frequency services and dynamic reduction of the largest loss can support a secure operational practice \cite{badesa_simultaneous_2019}.


%% file: classification.tex
\section{Classification of inertia estimation methods}
\label{sec:classification}
Cost-effective and secure operational decisions require accurate inertia estimates, which enable system operators to estimate the characteristics of the frequency response of the system. 
The frequency response in power systems is typically estimated based on the swing equation, as introduced in Eq. \eqref{eq:swing}, which relies upon the inertia in the system. Initial approaches to estimate inertia in power systems dominated by synchronous generator units assume the system inertia to be nearly constant, with an inertia constant of generation of approximately 5 seconds and an inertia constant of demand in the range of 1.6 - 1.8 seconds \cite{kelly_modelling_1994}.  
To consider the variability in the inertia level in power systems with reduced synchronous generator units and increased embedded units and uncertain and variable renewable energy resources, alternative inertia estimation methods have been developed that combine measurements of system variables, such as frequency and active power, and the physical model of the frequency profile. The system inertia is considered as a parameter to be estimated in a parameter estimation or system identification method. 
Inertia estimation methods in the literature can be categorized according to the time instant they are estimating the inertia for and according to the scope of the method. Table \ref{tab:estimation_literature} classifies the methods row-wise according to the time horizon and column-wise according to the scope. The following subsections will discuss this classification in more detail.

\input{./tables/classification.tex}

\subsection{Classification of inertia estimation methods according to the time horizon of interest}
\label{sec:time_horizon}
Inertia estimation methods can be categorized based on the time horizon of interest as offline post-mortem approaches, online real-time approaches and forecast approaches. Post-mortem inertia estimation methods are applied, for instance, in an analytical context to make an ex-post assessment of the inertia level in the system during large disturbance events. Real-time methods aim at making an instantaneous estimate of the inertia based on readily available measurements of system variables. Forecast approaches estimate the inertia to be expected in the future.

\subsubsection{Offline, post-mortem inertia estimation}
\label{sec:post_mortem}
Post-mortem, disturbance-based approaches to estimate total system inertia were the first attempts to improve the modelling-intensive approach in \cite{kelly_modelling_1994} to estimate the frequency response profile. Post-mortem approaches estimate the inertia after large frequency disturbing events have taken place for which the power imbalance related to the rate of change of frequency is accurately known. Inoue et al. were the first to introduce such a disturbance-based approach for inertia estimation that combines the swing equation with processed PMU measurements\footnote{PMUs can measure voltage and currents phasors in the system, as well as frequency, rate of frequency change, active power output of generators and power flows on transmission lines with measurement rates of 30 - 60 measurements per second \cite{phadke_synchronized_1993,shams_active_2019}. PMUs have the capability to directly monitor load feeders, but this is typically not the priority. The capability of PMUs to support inertia estimation depends on the noise and response delay on a step change \cite{wall_inertia_2014}. Also the locations of the PMUs impact the frequency measurement. Some methods use averaged frequency measurements of the PMUs to circumvent location specific frequency swings \cite{cao_switching_2016}.} of the frequency and the known size of the disturbance to estimate the total system inertia \cite{inoue_estimation_1997}. They estimate the inertia constant as:
\begin{equation}
  H= \frac{-\Delta P}{\left.\frac{d(\Delta f/f_0)}{dt}\right]_{t=0}}\label{eq:post-mortem}
\end{equation}
with $\Delta P$ the size of the loss and $\left.\frac{d(\Delta f/f_0)}{dt}\right]_{t=0}$ the rate of change of frequency at the starting time of the disturbance.\footnote{$\Delta f = f(t=0) - f_0$ with $f_0$ the nominal system frequency and t=0 the onset time of the disturbance.} This approach estimates the combined contribution of synchronous generators, embedded units and demand.
Most post-mortem, disturbance-based approaches ignore the damping and the short-term frequency response services in the imbalance $\Delta P$ applied in the swing equation, by assuming it to be zero in the short time period immediately after the disturbance \cite{chassin_estimation_2005,zhang_synchrophasor_2013}. Especially with the introduction of new, fast frequency response services, which reduce the time window of pure inertial response, this assumption may become challenging. Including activated frequency response services, such as governor response, and frequency and voltage dependent characteristics of the load in the modelling of the power imbalance is a way to deal with this issue, such as in   \cite{zografos_power_2018} and \cite{zografos_estimation_2017}. 
This approach also facilitates the simultaneous estimation of the system inertia constant and the size of the disturbance considering the frequency- and voltage-dependent dynamics using parameter estimation that relies upon solving a system of linear equations in \cite{zografos_power_2018} or an optimization in \cite{zografos_estimation_2017}.

Three aspects are specifically affecting the accuracy of post-mortem approaches.
First of all, the size of the loss should be accurately known. For this reason, not all events may be suitable for use in post-mortem inertia estimation. Ashton et al. mention the difficulty of estimating the inertia after events with multiple sequential outages of generators and suggest to focus on instantaneous events, which may be detected by their modified detrended fluctuation analysis \cite{ashton_application_2013}.
Second, the on-set of the event, i.e., t = 0, should be accurately determined. A recently suggested technique in the literature estimates the event on-set based on the second derivative of the ROCOF \cite{wang_fast_2020}. 
Third, the accuracy of the ROCOF calculation is affecting the accuracy of the inertia estimate. ROCOF has to be estimated on the relevant time interval, i.e., between the on-set of the event and the on-set of primary frequency response. The latter is determined by the time constant of the governors in traditional frequency response and is typically between 1-2 seconds following an infeed loss \cite{ashton_inertia_2015}.\footnote{This time constant may change in future power systems due to the integration of very fast frequency response services, such as the fast frequency response in Great Britain. These services do not react instantaneously, such as the inertia response, but within one second of a -0.5Hz change in frequency \cite{national_grid_system_2016,noauthor_enhanced_2019}.} The oscillatory component in the frequency signal produced by the synchronizing power between the generators may cause erroneous results for the ROCOF calculation. To overcome this issue, Inoue et al. fitted a polynomial to the frequency signal to extract the frequency signal and smooth the transients \cite{inoue_estimation_1997,chassin_estimation_2005,tavakoli_load_2012}, whereas Ashton et al. used a 0.5Hz low-pass filter to filter the frequency signal after finding the fitting of a polynomial to the signal insufficient \cite{ashton_inertia_2015}. Besides the oscillations, ROCOF calculations may be corrupted by noise, which is filtered using moving average filters \cite{inoue_estimation_1997} or modified detrended fluctuation analysis \cite{ashton_application_2013,ashton_inertia_2015} in the state-of-the-art techniques. Also the location of the frequency measurements relative to the in-feed loss is affecting the ROCOF calculation. For an identical event, a measurement from a weakly interconnected part of the system with low localized inertia results in a higher estimated value of the ROCOF compared to an estimate based on a measurement from an electrically strong part of the network with a relatively high inertia \cite{ashton_inertia_2015}. 

Important disadvantages of the post-mortem approaches are that increasing dynamic behaviour (small signal and transient) due to reduced inertia levels may hamper the accuracy of the inertia estimation and that the inertia can only be determined at discrete time instants, after a large disturbance event has occurred. As large disturbance events do not occur frequently, post-mortem approaches do not result in continuous inertia estimates. 
Inertia estimation methods with low temporal resolution are expected to become less reliable in the future due to the increasing variability of inertia in the system. Keeping a data base of accurate inertia estimates during large frequency events can support inertia estimation with higher temporal resolution.

\subsubsection{Online inertia estimation}
\label{sec:real_time}
In contrast to post-mortem estimates based on historical data of large disturbance events, real-time estimates use real-time measurements as an input for the inertia estimation. These real-time estimates may provide closer to real time information about the inertia in the system than post-mortem approaches. Three types of real-time inertia estimation methods with different temporal resolution can be distinguished: (i) estimating the total contribution from transmission-system-connected synchronous generator units as a lower bound on total inertia using information from the supervisory control and data acquisition (SCADA) system or energy management system (lower-bound methods) \cite{wilson_d2.3:_2018,orum_future_nodate}, (ii) close-to-real-time inertia estimation during disturbances based on PMU measurements (discrete methods) \cite{wilson_measuring_2019,wall_estimation_2012,wall_simultaneous_2014,guo_synchronous_2012,huang_generator_2013}, (iii) continuous inertia estimation based on PMU measurements or measurements with even higher sampling rates (continuous methods) \cite{tuttelberg_estimation_2018,berry_inertia_2019,zhang_online_2017}. 
Table \ref{tab:cat_RT} classifies the real-time approaches according to these three types of methods. The approaches are also classified according to the domain in which the inertia is estimated, i.e., using a time series model, a model in the Laplace domain or a model in the modal domain based on inter-area oscillations.

\begin{table}[htbp]
	\centering
	\begin{threeparttable}
	\caption{Categorization of real-time inertia estimation methods in lower bound, discrete and continuous methods that apply parameter estimation in time series, Laplace and modal domain models.}
	\label{tab:cat_RT}
	\begin{tabular}{r|ccc}
		\toprule
		& \multicolumn{3}{c}{Domain}\\
		&  Time series & Laplace & Modal  \\
		
		\midrule
		Lower bound & \cite{cao_switching_2016,orum_future_nodate} &  & \\
		Discrete &  \cite{wilson_measuring_2019,schiffer_online_2019,wall_estimation_2012,wall_simultaneous_2014,sun_-line_2019,huang_generator_2013,zhang_angle_2013} & & \cite{chavan_identification_2017,cai_inertia_2019,chakrabortty_measurement-based_2011,chow_estimation_2008,guo_synchronous_2012,panda_online_2019},\cite{guo_adaptive_2014}\tnote{1}   \\
		Continuous & \cite{berry_inertia_2019,panda_application_2019} & \cite{tuttelberg_estimation_2018,zhang_online_2017} & \cite{guo_adaptive_2014}\tnote{1}  \\
		
		\bottomrule
	\end{tabular}
	\begin{tablenotes}
		\small
		\item[1] Temporal resolution of the inertia estimates depends on the resolution of the estimation of the inter-area oscillations \cite{guo_adaptive_2014}. 
		\end{tablenotes}
	\end{threeparttable}
\end{table}

\paragraph{Total inertia from transmission-system-connected synchronous generation units}
Total inertia may be approximated by summing the inertia contribution of transmission-system-connected synchronous generators based on their online capacity and inertia constant. The inertia constant of a synchronous generator equals the kinetic energy in the rotating mass at rated speed of the generator expressed proportional to the power rating of the generator and represents the time in seconds a generator can provide rated power solely using the kinetic energy stored in the
rotating mass \cite{kundur_power_1994}. Accurate inertia constants of individual generators are typically not available to the transmission system operator \cite{winter_pushing_2015}. Also the equivalent inertia constant of virtual inertia resources is not readily available and variable over time. Research efforts have been focussing on estimating these equivalent inertia constants.\footnote{ These research efforts are indicated in the fourth column of Table \ref{tab:estimation_literature} and are discussed in more detail in Section \ref{sec:contributor_specific}.} On top of the inertia constant, this approach requires knowledge of the online status and the generation capacity of the generators. The online status of large transmission-system-connected generators can be monitored in the SCADA or EMS system, while their capacity is known to the system operator. 

Although the complexity of inertia estimation based on the online capacity of generators is very low, the resulting inertia estimate is not very accurate. First of all, the approach underestimates the total system inertia that is available, as only synchronous generators that are monitored by the transmission system operator are considered in the estimate, whereas small embedded units, such as combined heat and power plants, or small motor loads are not considered. Therefore, this result can be considered as a lower bound on the inertia, defining an upper limit on the ROCOF. 
Second, the inertia estimates resulting from lower bound inertia estimation are only available at discrete time instants. They rely upon the monitoring of energy management or SCADA systems. Although the SCADA system usually samples data at a relatively high frequency of typically 1 Hz, the standard practice is to store 10-minutes averaged values of the parameters characterising the operating and environmental conditions \cite{gonzalez_using_2019}. To estimate the inertia between two monitoring instants of the generators, a statistical model to estimate the change in inertia from synchronous generators based on measurements of frequency deviation with a higher temporal resolution can be applied \cite{cao_switching_2016}.

\paragraph{Discrete inertia estimation}
Inertia estimation methods are available that are able to estimate inertia close to real time during disturbances based on PMU measurements of frequency and estimates of power imbalance. Two main approaches can be distinguished in this group of methods: (i) Methods that directly work with the time series data and (ii) methods that work in the modal domain and estimate the inertia based on inter-area oscillations, as indicated in Table \ref{tab:cat_RT}. 
Similar to the offline disturbance-based approaches, the underlying model of these parameter estimation methods is the linearised swing equation, typically ignoring damping and primary and secondary frequency response. The methods assume to estimate the inertia within the very short period after the disturbance before these slower dynamics are in place. Schiffer et al. have included a generator governor model to capture the impact of primary frequency control \cite{schiffer_online_2019}. Alternatively, applying more detailed system models, including dynamic generator models and power flow equations, enable the estimation of the inertia based on voltage measurements \cite{petra_bayesian_2017}. However, this comes at an increased computational cost. Bayesian approaches for parameter estimation enable the quantification of the uncertainty on the point estimates of inertia caused amongst others by measurement noise \cite{petra_bayesian_2017}.

Real-time disturbance-based inertia estimation is challenging for two reasons. First of all, the online detection of an appropriate disturbance requires attention. Detection based on signals filtered using a moving-average filter is a frequently used technique \cite{wall_estimation_2012,wall_simultaneous_2014}. Second, accurate estimation of the size of the disturbance based on available PMU measurements is challenging. Approaches to deal with this challenge differ with the scope of the inertia estimation method. In methods that estimate the inertia constant of individual generators based on time series of PMU measurements it is frequently assumed that the mechanical power output of a generator equals the electrical power output at the previous time instant \cite{wall_estimation_2012,wall_simultaneous_2014}. This assumption is based on the slow response of a generators mechanical power output compared to the electrical power output and the fact that frequency control balances the mechanical and electrical power within the generator making them approximately equal before a disturbance \cite{sun_-line_2019}. Methods to estimate regional, on the contrary, approximate the power imbalance per independent frequency region as the change in power flow on the interconnecting lines during small disturbances \cite{wilson_measuring_2019}. The latter assumption originates from the inter-area power flow oscillations between independent frequency regions resulting from power imbalances.

The performance of disturbance-based inertia estimation methods depends on the size of the disturbance \cite{wilson_measuring_2019}. Originally, only outage events were considered as appropriate frequency events \cite{wall_simultaneous_2014}. However, also generator rescheduling at the hour causing frequent, recurring, and scheduled frequency variations have proven to be suitable for real-time inertia estimation \cite{schiffer_online_2019}. Nevertheless, the temporal resolution of the estimates is still limited to 30-60 minutes in these cases. At intermediate time instants, values for inertia have to be extrapolated based on the latest estimation \cite{cao_switching_2016}.

\paragraph{Continuous inertia estimation}
(Near-)continuous inertia estimation methods rely upon PMU measurements of the frequency or measurements with even higher sampling rates and continuous estimates of the power imbalance in the system. Based on these two sources of information they estimate inertia in the time domain, Laplace domain or modal domain, as indicated in Table \ref{tab:cat_RT}. (Near-)continuous methods have an increased temporal resolution compared to discrete inertia estimation methods. However, continuously estimating the power imbalance during normal system operation is hard, as power imbalances in normal operation are small compared to measurement noise and cannot be measured directly with state-of-the-art measurement equipment. Two approaches have been used in the literature to continuously determine the power imbalances in power systems during normal operation. A first approach estimates the power imbalances on a near-continuous basis (in a time scale of minutes or tens of minutes) based on PMU measurements. This method involves strong assumptions and is only valid if generator settings do not change \cite{tuttelberg_estimation_2018}. Second, inertia estimation methods based on microdisturbances may overcome the issue of poor estimation of power imbalance by injecting a known microdisturbance in the system. These microdisturbance may be optimized to maximize the signal to noise ratio \cite{zhang_online_2017} or the impact on the frequency measurements can be filtered with advanced signal processing \cite{berry_inertia_2019}.\footnote{The latter approach has been commercialized by Reactive Technologies and requires high-speed measurement units for very fast frequency measurements (no PMUs) \cite{berry_inertia_2019}.} 

The accuracy and reliability of continuous inertia estimation methods have not been proven so far. 
Guo et al have shown that the accuracy of their estimates reduce with larger errors in the measurements and that accurate initial guesses of the inertia constants are required to ensure the convergence of the algorithm, which are not straightforward to obtain \cite{guo_adaptive_2014}. 
Also the method of Tuttelberg et al. has shown deficiencies \cite{tuttelberg_estimation_2018}. The results of a case study indicate incorrect inertia estimates at some time instants. One of the reasons for the poor performance may be related to the approximation of the power imbalance in the system. The output of generators and interconnectors are considered to be constant over the measurement period and the measured changes in generation and interconnection flows compared to the start of the measurement period are attributed to power imbalances. Also microdisturbance methods have not been thoroughly validated. Validation of the method in \cite{berry_inertia_2019} is performed based on one week of inertia data provided by the British system operator, which are based on an empirical formula. Although the measurements swing around the empirical results, the empirical results do not correspond with the ground truth inertia. It is hard to attribute the swings of the measurements to the stochastic behaviour of inertia or to measurement noise, as values for the ground truth are not available and a reference case for calibration has not been presented. Moreover, it is important to verify the applicability of inertia estimation based on microdisturbances for large and not densely meshed systems  with high levels of inertia for which the single-centre-of-inertia approximation is not valid and where it is hard to distinguish small load disturbances from noise and natural attenuation of the perturbation \cite{yu_d2.1_2018}. Also potential power quality issues due to the inserted microdisturbances should be investigated \cite{yu_d2.1_2018}. 

\subsubsection{Estimation of expected inertia: Forecasting}
\label{sec:forecasting}
Due to the instantaneous nature of the inertial response, the system operator does not have time to take actions in real time to deal with insufficient levels of inertial energy available in the system. Therefore, the system operator should have an accurate estimate of the inertia that is expected to be available in the future to adequately modify the system's inertia level or potential largest loss in the system operator's decision-making process explained in Section \ref{sec:actions}.
Forecasts for different forecasting horizons can be used complementary to the frequency response requirement tables generated for different levels of demand, inertia and largest loss to identify when the system will potentially be at risk and prepare reductions of the largest infeed loss or actions to increase the inertia.
As contractual arrangements for generators to be online are made per 30 minutes, a 30 minute resolution of the forecasts is currently acceptable, although the increasing variability in generator output that comes with renewable energy sources may ask for a higher temporal resolution in the future.

The literature on models to forecast system inertia is limited to explanatory models that produce point forecasts of the system inertia by relating system inertia to system variables. These explanatory regression models rely upon forecasts of exogenous variables, such as expected generator status and load forecasts, to produce forecasts of inertia.\footnote{Abundant literature is available on load forecasting in power systems. An overview can be found in \cite{hong_probabilistic_2016,kuster_electrical_2017}.} A first approach to forecast system inertia only considers the contribution from transmission-system-connected synchronous generators that have the duty to report their expected status and has been reported by the independent system operator ERCOT \cite{du_forecast_2018}. However, the expected status of the generators is not generally available to all transmission system operators in day ahead. Moreover, a drawback of this model is that only considering transmission-system-connected synchronous generators' contribution results in an underestimation of the total system inertia as the contribution from demand has not been considered. Other system operators estimate the future inertia contribution from embedded units and demand based on the demand forecast multiplied with an empirical constant \cite{wilson_measuring_2019}. The resulting structure of the forecast model for the inertial energy $\hat{E}^{I}_{t+k\vert t}$ can be represented as:
\begin{equation}
\hat{E}^{I}_{t+k\vert t} = \sum_i \hat{K}_{t+k \vert t, i} \cdot H_i \cdot P^{G,C}_{i} + a \cdot \hat{P}^{D}_{t +k \vert t} \label{eq:forecast}
\end{equation}
with $\hat{K}_{t+k \vert t, i}$ the reported status of the synchronous generator unit $i$ at time $t$ for time $t+k$, $H_i$ its inertia constant [s], $P^{G,C}_{i}$ its generating capacity [MW] and $\hat{P}^{D}_{t +k \vert t}$ the forecast of the system demand at time $t$ for time $t+k$ [MW].
While originally a single constant regression constant $a$ has been used for the system as a whole, linear regression analysis on regional inertia estimates has been applied to derive spatially differentiated empirical constants \cite{wilson_measuring_2019}. To build these regression models, historical data of the inertia in the system and the corresponding system state are required. The accuracy of the forecasts will be determined by the accuracy and resolution of the historical inertia estimates. 

Point forecast models relying upon the reported status of synchronous generator units are expected to decrease in accuracy. Due to the increased penetration of variable and uncertain renewable energy sources with zero marginal cost, the market dynamics are expected to increase and the day-ahead reported statuses are expected to be less accurate and more uncertain, which hampers the forecasting of the available inertia. This effect has been shown in the ERCOT system, where inertia forecasts based on week-ahead generator reports typically overestimate the real-time inertia during periods with low energy prices \cite{du_forecast_2018}. Two measures can be taken to deal with this challenge. First of all, an improved point forecast model of the status of the synchronous generators may reduce the point forecast error. A first approach is to model the energy trading to estimate the generators' statuses. However, modelling energy trading is not straightforward, especially not if quantitative results are required \cite{weron_electricity_2014}. Moreover, transmission system operators in power systems with independent market operators do not have access to the bids and offers in the energy markets and are only informed by the outcome of the market at gate-closure time close to real time, e.g., one hour ahead of real time in Great Britain \cite{noauthor_balancing_nodate}. Therefore, it might be more beneficial from a system operator's perspective to identify relationships between the generators' online capacity and other system variables impacting the energy trading that are directly available to the system operator, such as historical data of the status of the different generators, historical data and forecasts of total transmission system demand, forecasts of wind and solar generation, interconnector flows, electricity prices, etc. Second, while current practice focuses on point forecasts, modelling the uncertainty on the forecasts enables the system operator to assess the risk of ROCOF and underfrequency relay tripping. Probabilistic forecast models are a useful tool in this respect, as they represent the conditional distribution of the future inertial energy for different forecast horizons conditional upon the system state, for a variety of system states. The system state is characterized by a set of exogenous variables, such as demand, wind production, day-ahead electricity price and forecasts of the respective system variables. The linear regression model in state-of-the-art approaches provides a framework to estimate or predictive distribution of the inertia forecasts, but this is based on a set of strict assumptions, i.e., (i) the errors are statistically independent, (ii) the errors have a constant variance and (iii) follow a normal distribution. Alternative parametric, semi-parametric and non-parametric probabilistic forecast models should be developed for inertia forecasting.\footnote{Probabilistic forecasting models have already been applied extensively in other energy forecasting contexts, such as load forecast, price forecasting and wind forecasting \cite{hong_probabilistic_2016,nowotarski_recent_2018,zhang_review_2014}.} The development of accurate probabilistic forecast models requires a large data set. However, current inertia estimation methods provide accurate estimates during discrete events resulting in a data set too small for accurate forecast modelling or near-continuous estimates that are inaccurate or of which the accuracy has not been proven.

Traditional explanatory inertia forecasting models as the one introduced in Eq. \eqref{eq:forecast} rely upon the assumptions that a linear, additive and stationary relation holds between system inertia and the exogenous variables, i.e., total system demand and the inertia contribution from synchronous generators. 
To avoid the assumption of linearity and additivity, the feasibility of non-linear, artificial neural network (ANN) models as explanatory models for inertia forecasting has been tested in \cite{wilson_d2.3:_2018}. The presented ANN models use total system demand, total generation from power-electronic interfaced generation and total generation from synchronous generators as input features to model the system inertia. General statements about the accuracy of the method cannot be made as the report only mentions the application of the approach in a simplified simulation context, which does not capture the actual stochastic behaviour of the inertia in the system \cite{wilson_d2.3:_2018}.

\subsection{Scope of inertia estimation methods}
\label{sec:scope}
Inertia estimation methods may also be categorized based on their spatial resolution. Some methods are able to estimate the total system inertia or regional inertia, which makes them different in terms of their geographical scope. Other methods focus on the estimation of different contributors to the inertia, such as synchronous generators, demand or virtual inertia resources.

\subsubsection{Geographical scope}
\label{sec:geographical_scope}
Historical approaches to estimate the system inertia focussed on calculating a single value for the total system inertia based on the concept of centre of inertia, which is determined by the average system frequency \cite{inoue_estimation_1997,chassin_estimation_2005,zografos_power_2018,zografos_estimation_2017}. This value of system inertia captures the contribution from generation, demand and virtual inertia resources, when present. Assuming a single value for inertia in the system is acceptable in strong transmission systems with limited variations of the frequency between locations. Lower inertia levels will, however, increase the risk for angle swings and oscillations between areas \cite{kundur_power_1994}. 

To improve the inertia estimates, recent efforts have been focussing on providing zonal estimates of the inertia or spatial inertia profiles. Zonal inertia estimates are more flexible than system inertia estimates, as aggregating the zonal inertia contributions gives the total system inertia, while the zonal or individual contributions cannot be derived from the system estimate. These zonal inertia estimates may be used by the system operator in procuring very fast frequency response services and in enhanced and coordinated frequency control methods \cite{noauthor_enhanced_2019}. Different approaches to divide the system in zones have been used in the literature: (i) division around the constraint boundaries of the network under analysis \cite{ashton_inertia_2015} or (ii) in zones in which angles and frequency are closely coupled \cite{wilson_measuring_2019,chavan_identification_2017,chakrabortty_measurement-based_2011}. The latter is preferable from a physical perspective.

\subsubsection{Contributors to the system inertia}
\label{sec:contributor_specific}
From a system operator perspective, it might be useful to get insight in the different contributors to the aggregated system inertia in a zone or in the total system. When total inertia levels are decreasing, the contribution of unmonitored contributors to the system inertia, such as demand and embedded units, will become more important and accurate estimates of these contributions are required. Forecasts of the inertia contribution of embedded units and demand may be used in advanced unit commitment models that simultaneously schedule energy and frequency response services \cite{badesa_optimal_2019} or can determine the volume of inertia to buy on an inertia market.

Directly estimating the inertia contribution from demand and embedded units is challenging, which is also reflected in the limited research efforts in Table \ref{tab:estimation_literature}. Initially, fixed values for the contribution of load to the system inertia were used (e.g., 20\% of the total inertia in Great Britain \cite{ashton_inertia_2015}) or the load contribution was just omitted \cite{concordia_load_1982}. Ashton et al. indicated in 2015 that the contribution of demand and embedded units, such as distribution connected synchronous generator units, to total system inertia varies between 8\% and 25\%, with an average of 18.18\% \cite{ashton_inertia_2015}. The inertia contribution from embedded units and load has been estimated by subtracting the contribution from synchronous generator units from the total system inertia estimate during large disturbance events \cite{tavakoli_load_2012,bian_demand_2018}. Estimating the contribution from synchronous generator units requires information about their online capacity\footnote{If information about the generators' online capacity is not available, Bian et al. have proposed a method to estimate the generator contribution based on the produced power by fuel type \cite{bian_demand_2018}.} and their individual inertia constants. Originally, synchronous generators' inertia constants were estimated using the load ramp test, the probing test and the transient test \cite{ariff_estimating_2014,tsai_line_1995,karrari_identification_2004,noauthor_IEEE_2010}. Alternatively, dynamic online estimation based on PMU measurements has been proposed to estimate the inertia constants of synchronous generator units or virtual inertia resources, as indicated in the third and fourth column of Table \ref{tab:estimation_literature}. In these methods, inertia is estimated as a parameter in a dynamic frequency responsive model, matching the power output of the model with the power output from the measurements during the disturbance event. The comparison to derive the inertia constants can be done manually \cite{littler_measurement-based_2005} or using a grey-box identification method that optimizes the parameters of a dynamic model of the frequency response of a generator unit by minimizing a weighted mean squared error between the power output from the dynamic model for the given frequency measurements related to the disturbance with the measured power output of the generator unit \cite{tavakoli_load_2012}.\footnote{Different simulation models should be built for different types of generator units.} Alternatively, a two stage validation and calibration process has been suggested to validate the dynamic model accuracy of a generator based on PMU measurements and update the model parameters using an extended Kalman filter in case of model deficiencies \cite{huang_generator_2013}. One of the dynamic model parameters to be estimated in this approach is the inertia constant of the generator.

Based on the limited data set of estimates of the inertia contribution from embedded units and demand during large disturbance events, research efforts have developed explanatory models that relate the inertia contribution from demand and embedded units to the total system demand  \cite{bian_demand_2018,tavakoli_load_2012}. These models have predictive capabilities. However, due to the limited size of the data set, only simple model structures with a limited number of parameters to be trained, such as linear regression models, have been used to avoid overfitting \cite{bian_demand_2018}. 
The temporal resolution of the estimates of the inertia contribution of demand and embedded units, and therefore the size of the available data set, can be increased by using an estimation method with higher temporal resolution to estimate the total system or zonal inertia.\footnote{An overview of different methods for total system or zonal inertia estimates with different temporal resolution with their characteristics and shortcomings has been given in Section \ref{sec:time_horizon}.}


%% file: tables/classification.tex
  \begin{sidewaystable}[htbp]
    \begin{threeparttable}
    \caption{Overview and classification of inertia estimation methods according to time horizon and scope}
    \label{tab:estimation_literature}
    \centering
    \begin{tabular}{r|C{2.8cm}C{2cm}C{3.6cm}C{2.8cm}C{2.8cm}C{2.8cm}}
      \toprule
       & \multicolumn{6}{c}{Scope}\\
         &  System & Zone & Synchronous generation & Embedded generation & Clustered generators/Node & Demand\\
         
      \midrule
      Offline, post-mortem & \cite{ashton_inertia_2015,inoue_estimation_1997, chassin_estimation_2005,zografos_power_2018, ashton_application_2013,zografos_estimation_2017} & \cite{chavan_identification_2017} & \cite{tavakoli_load_2012} & \cite{littler_measurement-based_2005,zhang_synchrophasor_2013} & & \cite{tavakoli_load_2012,bian_demand_2018}\\
      \midrule
        Online, discrete& \cite{schiffer_online_2019,cai_inertia_2019,wall_simultaneous_2014} & \cite{wilson_measuring_2019,chakrabortty_measurement-based_2011,chow_estimation_2008} & \cite{wall_estimation_2012,sun_-line_2019,huang_generator_2013,kalsi_calibrating_2011,guo_synchronous_2012} & \cite{petra_bayesian_2017} & \cite{panda_online_2019} &\\
       \midrule
      Online, continuous & \cite{berry_inertia_2019} &  \cite{tuttelberg_estimation_2018} & \cite{panda_application_2019,guo_adaptive_2014} & & \cite{zhang_online_2017} & \\
      \midrule
         Forecast & \cite{du_forecast_2018, noauthor_ge_nodate} & \cite{wilson_d2.3:_2018} & & & &\\
         
         \bottomrule
    \end{tabular}
    \begin{tablenotes}
      \small
      \item References that present a method that is able to estimate the inertia for different scopes are only mentioned once in the table, i.e., in the column of the smallest scope it is applicable to.
    \end{tablenotes}
    \end{threeparttable}
    \end{sidewaystable}

%% file: conclusion2.tex
\section{Conclusion and the future of inertia estimation methods}
\label{sec:conclusion}
Secure and cost-effective operation of low-inertia power systems requires accurate and reliable estimates and forecasts of inertia in the system on a close-to-continuous and regional basis, considering contributions from generation, demand and virtual inertia resources. The state of the art in inertia estimation methods has evolved from delivering post-mortem, total system inertia estimates after discrete large power imbalance events towards close-to-continuous real-time inertia estimates with the possibility to make regional estimates, which is crucial in power systems consisting of strong regions interconnected by long links.  
These near-continuous inertia estimates enable monitoring of the system's frequency response profiles with a high temporal resolution, which is crucial to deal with the operational challenges induced in power systems with decreasing inertia levels. 
Although important steps have been taken towards near-continuous inertia estimates, the accuracy and reliability of continuous inertia estimation techniques have not been adequately proven yet due to limited ground truth knowledge about the inertia in the system. Recommended directions for further work and future research are (i) the testing and validation of near-continuous inertia estimation methods to ensure that the techniques do not impact the stability of the system and to obtain more accurate estimates, (ii) getting more insight in the contributors to the system inertia in different operating states, especially the contribution of embedded units that are not observable and uncontrollable for the system operator, and (iii) the development of models to forecast future levels of total system inertia, zonal inertia or the inertia contribution from individual resources, such as demand and unmonitored embedded units. 

First of all, the proposed near-continuous inertia estimation methods and technologies should be validated and calibrated against situations during which the inertia is accurately known, such as the testing of a single generator or a large frequency event. Moreover, further work is needed to verify the impact on the stability and security of the power system of the small power injections in microdisturbance methods. Based on the results of these analyses the presented methods may be further improved. 

Second, having insight in the contribution of demand and embedded units to inertia and its relation to other system variables is important in power systems with decreasing inertia levels to inform power system operators' decision making process about the unmonitored inertia in the system in real time and in the future. Estimates of the inertia contribution of demand and embedded units rely upon accurate estimates of the total inertia from which the inertia contribution of monitored, transmission-system-connected generators is subtracted. Due to the limited availability of reliable and accurate total system inertia estimates with high temporal resolution, estimation of the contribution of demand and embedded units currently relies upon inertia estimates during large disturbance events. 

Third, the field of inertia forecasting has ample opportunities for progression. The state of the art on inertia forecasting models is limited to explanatory, linear regression models with a limited number of exogenous variables and focuses on providing point forecasts. However, the system operator would benefit from probabilistic inertia forecasting models to assess the risk of ROCOF and underfrequency relay tripping and to inform decision making.  
Once reliable and accurate inertia estimates of the total system or a zone as well as of the different contributors are available  with a high temporal resolution, these data can be used to develop probabilistic forecasting models. The resulting probabilistic forecasts of inertia may inform system operator's decisions on the modification of the inertia in the system or the potential largest loss, frequency response procurement or the volume of inertia to buy on a market for inertia. 
Alternatively, one can look into the benefits of direct forecasts of predictive distributions of system variables, such as rate of change of frequency and frequency deviation, rather than the system parameter of inertia to estimate the probability of potential issues in the frequency response. Similar approaches of integrated forecasting have been suggested in the context of determining market bids of power plants with uncertain production \cite{carriere_integrated_2019}.
